\documentstyle{ioplppt}

\newcommand{\de}{\partial}

\def\bt#1{\mbox{\bf #1}}
\def\ct#1{\mbox{\cal #1}}

\newcommand{\ds}{\displaystyle}

\renewcommand{\>}{\rangle}

\renewcommand{\bs}{\backslash}

\newcommand{\Max}{\mbox{\scriptsize max}}
\newcommand{\Min}{\mbox{\scriptsize min}}
\newcommand{\OBC}{\mbox{\scriptsize OBC}}
\newcommand{\CBC}{\mbox{\scriptsize CBC}}
\newcommand{\Inf}{\mbox{\scriptsize inf}}
\newcommand{\Sup}{\mbox{\scriptsize sup}}
\renewcommand{\tr}{\mbox{\scriptsize tr}}
\newcommand{\bound}{\mbox{\scriptsize bound}}
\newcommand{\proof}{{\em Proof.\ }}

\newtheorem{definition}{Definition}[section]
\newtheorem{lemma}{Lemma}[section]
\newtheorem{proposition}{Proposition}[section]
\newtheorem{theorem}{Theorem}[section]

\begin{document}
\jl{1}

\title{On the Thermodynamic Limit in Random Resistors Networks}[
Random Resistors Networks]
\author{Francesco Guerra\ftnote{1}{e-mail: guerra@roma1.infn.it} 
and Mauro Talevi\ftnote{2}{e-mail: talevi@roma1.infn.it}}
\address{Dipartimento di Fisica, Universit\`a di Roma ``La Sapienza" 
and INFN, Sezione di Roma, Piazzale Aldo Moro 2, 00185 Roma, Italy}
\begin{abstract}
We study a random resistors network model on a euclidean geometry
$\bt{Z}^d$.  We formulate the model in terms of a variational 
principle and show that, under appropriate boundary conditions, the
thermodynamic limit of the dissipation per unit volume is finite almost
surely and in the mean.  Moreover, we show that for a particular thermodynamic
limit the result is also independent of the boundary conditions.
\end{abstract}

\pacs{1315, 9440}

\maketitle

\section{Introduction}
 
In this paper we study a model of 
random resistors networks (RRN) on a euclidean geometry $\bt{Z}^d$.
RRN are examples of disordered statistical mechanical systems
which have been widely considered in literature in the context of
percolation theory \cite{percolation}, with different 
lattice geometries and different probability distributions for the resistors
\cite{BE}--\cite{Golden2}.  

The problem we address is the behaviour of the RRN model  
in the thermodynamic limit, concentrating in particular 
on the role of the boundary conditions.  
We shall restrict ourselves to the case in which the resistors are 
independent and identically distributed positive random variables.
We do not specify any distribution function but assume it to be 
smooth enough to have a finite expectation value.  
The physical observable we consider is the 
dissipation per unit volume, which is related to the total 
conductance of the network (cf.~equation (\ref{eq:WvsC})).
We show that, under appropriate boundary 
conditions (which we call {\em closed boundary conditions} (CBC) and will be 
specified in section~\ref{sec:model}), 
the thermodynamic limit of the dissipation is finite,
both in the mean and almost surely.   More precisely, the main results of
this paper are summarized by the following theorem.

\begin{theorem} 
\label{th:recap}
If the conductances of the network have a finite expectation value $\<C\>$,
then the limit of the dissipation per unit volume 
$\ds{\frac{W^{\CBC}_{LA}}{LA}}$ on a rectangular\footnote{We consider the
case of $d=2$, being the extension to the case of an arbitrary $d$ trivial.}
RRN of dimensions $(L,A)$,
\begin{equation}
\lim_{L,A\to\infty}\frac{W^{\CBC}_{LA}}{LA}=v^2_0\bar c^{\CBC}
\end{equation}
exists in the mean and a.s. and is finite, independently of the order of the 
limits on $L$ and $A$, where $v_0$ is a real positive number,
\begin{equation}
v^2_0\bar c^{\CBC}=\lim_{L,A\to\infty}\frac{\bt{E}(W^{\CBC}_{LA})}{LA}.
\end{equation}
and CBC denotes closed boundary conditions.  
Moreover, if we let $A\to\infty$ before $L\to\infty$, then 
\begin{equation}
\lim_{L\to\infty}\lim_{A\to\infty}\frac{W_{LA}}{LA}=
v^2_0\bar c^{\CBC}
\end{equation}
exists in the mean and a.s. and is finite, independently of the boundary 
conditions.
\end{theorem}

This paper is organized as follows.
In section~\ref{sec:model} we define the RRN model based on the classical laws 
of Ohm and Kirchoff.  
In section~\ref{sec:vp} we show how to formulate the model based on a 
variational principle and in section~\ref{sec:tlr} we use this principle 
to study the thermodynamic limit of the dissipation per unit volume.

\section{The model}
\label{sec:model}

We consider a RRN model with an euclidean geometry $\bt{Z}^d$ and denote with
$R_{nn'}$ the resistors on the links $(n,n')$ of $\bt{Z}^d$, which are
taken to be independent and identically distributed random variables.   
The model is defined on a finite set $\Lambda\subset \bt{Z}^d$.  We fix 
a direction, e.g. the direction 1, along which a potential difference is
established between the $(d-1)$-dimensional hypersurfaces which are 
perpendicular to the direction 1.  Equally valid RRN can be obtained
with current generators instead of potential generators \cite{BCES}.

Boundary conditions (BC) play a fundamental role in all statistical system
\cite{GRS,Ruelle}.  
The boundary of the finite region $\Lambda\subset\bt{Z}^d$ is defined as 
$\de\Lambda=\{n\in\Lambda:\ \exists\ n',\ n'\not\in\Lambda,\ |n-n'|=1 \}$.
The BC we use are imposed by fixing the potential on $\de\Lambda$
such that
\begin{equation}
V_{n}=n_1v_0,\qquad n=\{n_1,\ldots,n_d\}\in \de\Lambda,
\label{eq:BC}
\end{equation}
where $v_0$ is a real positive number.
If we indicate the set of links contained in $\Lambda$ with
$L_{\Lambda}=\{(n,n')\subset\Lambda:\ |n-n'|=1\}$,
we can assign on $L_{\Lambda}$ the {\em conductance field}
$C:L_{\Lambda} \to \bt{R}$, 
where $C_{nn'}$ are non negative, independent and identically distributed 
random variables, independent of the link orientation, i.e.
$C_{nn'}=C_{n'n},\ (n,n')\in L_{\Lambda}$.
We denote with $(\Omega,\ct{F},\bt{P})$ the probability space on which 
the variables $C_{nn'}$ live, and with $\bt{E}(\cdot)$ the expectation value
respect to the measure $\bt{P}$.  We use the notation 
$\bt{E}(C^k_{nn'})=\<C^k\>,\ k=1,2,\ldots,\ (n,n')\in L_{\Lambda}$,
for the expectation value of the $k$-th power of the variables $C_{nn'}$.

The {\em potential field} $V:\Lambda\to\bt{R}$ 
is related to the conductance field $C$ by Ohm's law
\begin{equation}
(V_n-V_{n'})C_{nn'}=I_{nn'},\qquad  (n,n')\in L_{\Lambda},
\label{eq:Ohm}
\end{equation}
where $I_{nn'}$ is the current passing though the link $(n,n')$.  
Denoting with $\bar\Lambda\subset\Lambda$ the set of points of $\Lambda$ 
on which the potential field $V$ is fixed to a value $\bar V$, 
Kirchoff's First Law is given by  
\begin{equation}
\sum_{\{n'\in\Lambda:\ |n'-n|=1\}}I_{nn'}=
\sum_{\{n'\in\Lambda:\ |n'-n|=1\}}(V_n-V_{n'})C_{nn'}=0,
                     \ n\in \Lambda\bs\bar\Lambda.
\label{eq:Kirchoff}
\end{equation}
Thus, by Ohm's and Kirchoff's laws, on the sites
$n\in\Lambda\bs\bar\Lambda$ there is a well-defined potential 
\begin{equation}
V_n=\frac{\sum_{\{n'\in\Lambda:\ |n'-n|=1\}} V_{n'} C_{nn'}}{
      \sum_{\{n'\in\Lambda:\ |n'-n|=1\}} C_{nn'}},
                     \qquad n\in \Lambda\bs\bar\Lambda,
\label{eq:wa}
\end{equation}
Note that Kirchoff's and Ohm's laws are relations
valid for each $\omega\in\Omega$.
The following proposition will be useful in the following.
\begin{proposition}[Maximum Principle]
\label{prop:max}
Let $\bar V_n,\ n\in\bar\Lambda$ the values of the fixed potential
and $V_{\Max}$, $V_{\Min}$ its maximum and minimum values,
$V_{\Min}\leq \bar V_n\leq V_{\Max},\ n\in\bar\Lambda$.
Then, for each $\omega\in\Omega$, 
$V_{\Min}\leq V_n\leq V_{\Max},\ n\in \Lambda\bs\bar\Lambda$.
\end{proposition}
\proof
Let us show e.g. the inequality $V_n\leq V_{\Max}$.  
By equation~(\ref{eq:wa}), the potentials 
$V_n,\ n\in \Lambda\bs\bar\Lambda$, are given by a weighted average
of $V_{n'}$ on the nearest-neighbor sites $n'\in\Lambda$, i.e. 
for every $\omega\in\Omega$
\begin{equation}
V_n\leq \max_{\{n':\ |n'-n|=1\}}V_{n'},\qquad n\in \Lambda\bs\bar\Lambda,
                                        \ n'\in \Lambda.
\label{eq:estimate}
\end{equation}
If there exists a point $\bar n\in \Lambda\bs\bar\Lambda$ such that  
$V_{\bar n}>V_{\max}$, then we would have $V_{\bar n}>\bar V_{n'}$, 
for every $n'\in\bar\Lambda$.  On the other hand, applying
repeatedly equation~(\ref{eq:estimate}) we can find a $n'\in\bar\Lambda$ 
such that $V_{\bar n}\leq \bar V_{n'}$, hence the contradiction.

\section{Variational principle}
\label{sec:vp}

In this section we show that Kirchoff's law, cf.~equation~(\ref{eq:Kirchoff}),
can be obtained by a variational principle.
In particular, the potential field $V$ determined by Kirchoff's law 
minimize the dissipation per unit volume by Joule effect 
by the network in the region $\Lambda$
\begin{equation}
w_{\Lambda}(C,\bar V)=\frac{W_{\Lambda}(C,\bar V)}{|\Lambda|}
=\frac{1}{|\Lambda|}\sum_{(n,n')\in L_{\Lambda}}(V_n-V_{n'})^2C_{nn'},
\label{eq:dissipation}
\end{equation}
where $\bar V$ denotes a field of fixed values for the potential on 
$\bar\Lambda$ and the sum is taken over all the links $(n,n')$ in the region
$\Lambda$\footnote{We stress that notation $(n,n')\in L_{\Lambda}$ should be 
intended as a sum over the number of links and not simply as a sum over 
nearest-neighbor points $n$ and $n'$.  The latter sum differs from the former 
by a factor of 2.}.  We introduce the set of the potential fields
which coincide with $\bar V$ on $\bar\Lambda$,
$\Phi(\bar\Lambda,\bar V)=\{\phi:\ \phi_n=\bar V_n,\ n\in\bar\Lambda\}$,  
and the {\em dissipation functional} 
$\varphi_{\Lambda}:\Phi(\bar\Lambda,\bar V)\to\bt{R}$
\begin{equation}
\varphi_{\Lambda}(C,\phi)=\sum_{(n,n')\in L_{\Lambda}}
                          (\phi_n-\phi_{n'})^2C_{nn'}.
\end{equation}
\begin{definition}
\label{def:field-variations}
Given a field $\phi\in\Phi(\bar\Lambda,\bar V)$, the function
$ U^{\epsilon}_n: \Lambda\times [-1,1]\to \bt{R} $
is called the variation of the field $\phi$ in $\Phi(\bar\Lambda,\bar V)$ if
\begin{enumerate}
\item $U^{\epsilon=0}_n=\phi_n$, for every $n\in\Lambda$, 
\item $U^{\epsilon}_n=\bar V_n$, for every $n\in\bar\Lambda$
                                 and $\epsilon\in [-1,1]$,
\item $U^{\epsilon}\in C^{\infty}(\Lambda\times [-1,1])$.
\end{enumerate}
We denote with $\ct{V}(\phi)$ the set of variations of $\phi$ 
in $\Phi(\bar\Lambda,\bar V)$.
\end{definition}

\begin{definition}
\label{def:stationarity}
$V\in\Phi(\bar\Lambda,\bar V)$ is called a stationary point for 
$\varphi_{\Lambda}(C,\phi)$ on
$\Phi(\bar\Lambda,\bar V)$ if
the function $\varphi_{\Lambda}(C,U^{\epsilon}):[-1,1]\to \bt{R}$
has a stationary point in $\epsilon=0$ for every $U^{\epsilon}\in \ct{V}(V)$.
\end{definition}

We can now prove the main result of this section:

\begin{theorem}[Least Dissipation Principle]
\label{th:ldp}
The potential field $V\in\Phi(\bar\Lambda,\bar V)$ determined by
Kirchoff's law is a minimum point for the functional
$\varphi_{\Lambda}(C,\phi)$, i.e.
\begin{equation}
W_{\Lambda}(C,\bar V)=\min_{\{\phi:\ \phi_n=\bar V_n,\ n\in\bar\Lambda\}}
                     \varphi_{\Lambda}(C,\phi).
\end{equation}
\end{theorem}
\proof
We want to show that the potential field $V\in\Phi(\bar\Lambda,\bar V)$ 
determined by Kirchoff's law is both a stationary point and a minimum point
for the dissipation functional $\varphi_{\Lambda}(C,\phi)$.

The stationarity is proven by noting that 
for every $U^{\epsilon}\in\ct{V}(V),\ V\in\Phi(\bar\Lambda,\bar V)$,
\begin{equation}
\left.\frac{d}{d\epsilon}\varphi_{\Lambda}(C,U^{\epsilon})\right|_{\epsilon=0}
          =2 \sum_{n\in\Lambda\bs\bar\Lambda}
       \left(\sum_{\{n'\in\Lambda:\ |n'-n|=1\}}(V_n-V_{n'})C_{nn'}\right)Z_n,
\end{equation}
having used the fact that, by definition \ref{def:field-variations},
$Z_n=\de U^{\epsilon}_n/\de\epsilon |_{\epsilon=0}=0$
for every $n\in\bar\Lambda$.
Then, by Kirchoff's law (\ref{eq:Kirchoff}) and 
the arbitrariness of the $Z_n,\ n\in \Lambda\bs\bar\Lambda$, 
consequence of the arbitrariness of the $U^{\epsilon}\in\ct{V}(V)$, 
definition~\ref{def:stationarity} implies that 
$V\in\Phi(\bar\Lambda,\bar V)$ is a stationary point
for $\varphi_{\Lambda}(C,\phi)$. 

To prove that the potential field $V\in\Phi(\bar\Lambda,\bar V)$ 
is a minimum point, we need to show that 
$\varphi_{\Lambda}(C,U^{\epsilon})\geq \varphi_{\Lambda}(C,V)$,
for every $U^{\epsilon}\in\ct{V}(V)$ and every $\epsilon\in [-1,1]$.
Putting $U^{\epsilon}_n=V_n+Z^{\epsilon}_n$ we obtain
\begin{eqnarray*}
\varphi_{\Lambda}(C,U^{\epsilon})-\varphi_{\Lambda}(C,V)
         &\geq&2\sum_{(n,n')\in L_{\Lambda}}(V_n-V_{n'})
               (Z^{\epsilon}_n-Z^{\epsilon}_{n'})C_{nn'} \\
           &=&2\sum_{n\in\Lambda}\left(\sum_{\{n'\in\Lambda:\ |n'-n|=1\}}
              (V_n-V_{n'})C_{nn'}\right) Z^{\epsilon}_n.
\end{eqnarray*}
Then, using Kirchoff's law (\ref{eq:Kirchoff}), we get
$\varphi_{\Lambda}(C,U^{\epsilon})-\varphi_{\Lambda}(C,V)\geq 0$,
for each $Z^{\epsilon}_n$, i.e. for every $\epsilon\in [-1,1]$.

\section{Thermodynamic limit}
\label{sec:tlr}

In this section, we study the thermodynamic limit of the 
dissipation density $\ds{\frac{W_{\Lambda}}{|\Lambda|}}$, as
 the volume $|\Lambda|$ of the region $\Lambda\subset\bt{Z}^d$ goes to infinity.
We parametrize the region $\Lambda\subset \bt{Z}^d$ with a
rectangle of dimensions $a_1,\ldots,a_d$ and then take the limit 
$a_i\to\infty,\ i=1,\ldots,d$.  
For notational simplicity we shall consider the case with $d=2$, as the
results are trivially extendable to arbitrary $d$.
We parametrize the rectangle by a pair of integers $(L,A)$, where 
$L$ is the number of links in the longitudinal direction
and $A$ is the number of sites in the transverse direction.

We denote the dissipation density in two dimensions with\footnote{It must
be stressed that, even if the notation might be slightly misleading, 
the dissipation is a function of the region $\Lambda(L,A)$ of dimensions 
$(L,A)$ and not only of the dimensions $(L,A)$.  It would be more appropriate
to write $W_{\Lambda(L,A)}$, but when not strictly necessary for the 
comprehension, we shall use the abbreviated form.}
\begin{equation}
\frac{W_{LA}}{LA}=\frac{1}{LA}\sum_{(n,n')\in L_{\Lambda}}(V_n-V_{n'})^2C_{nn'}.
\label{eq:w_LA}
\end{equation}
Let us establish the potential diffence along the longitudinal direction. 
We shall consider the case in which no boundary conditions are imposed along the
longitudinal direction ({\em open BC}), and  the case in
which the boundary conditions are imposed following equation~(\ref{eq:BC})
({\em closed BC}).

Let us note that, being the potential difference
proportional to the longitudinal dimension, the dissipation density can be
written in terms of the total conductance $C_{LA}$ of the network,
\begin{equation}
\frac{W_{LA}}{LA}=\frac{V^2_LC_{LA}}{LA}=v^2_0\frac{L}{A}C_{LA}.  
\label{eq:WvsC}
\end{equation}
Indeed, the authors of \cite{BCES}
study the thermodynamic limit of the r.h.s. of equation~(\ref{eq:WvsC}).

\subsection{Preliminary lemmas}

In this section we apply the variational principle of section~\ref{sec:vp}
to derive a few properties for the dissipation $W_{LA}$.  In particular,
we show that $W_{LA}$ is a subadditive and superadditive sequence of
random values, respect to $L$ and $A$, depending on the BC.
We refer the reader to \cite{BCES,Kingman,Ruelle79}, for the definition 
of subadditivite and superadditivite sequences, 
while the main theorems we use in this paper are reported in 
\ref{app:sss}.

In the following OBC and CBC denote the open and closed BC
respectively, while $\#$ both BC.

\begin{lemma}
\label{lemma:OBC<CBC}
For every $\omega\in\Omega$, $W^{\OBC}_{LA}(\omega)\leq W^{\CBC}_{LA}(\omega)$.
\end{lemma}
\proof
By the Least Dissipation Principle, theorem \ref{th:ldp}, 
the total dissipation $W(C,\bar V)$ is obtained minimizing, for every 
$\omega\in\Omega$, the functional $\varphi(C,\phi)$ respect to all  
test fields $\phi$, with the constraint that 
$\phi_n=\bar V_n,\ \forall n\in\bar\Lambda$.  Since by imposing more constraints
on $\phi$ we restrict the space on which they can vary, we have that
for every $\omega\in\Omega$ the minimum of $\varphi(C,\phi)$ 
on the reduced space will be greater or equal to the one on the 
space with less constraints.  Hence, the thesis follows by considering the
closed BC as a greater number of constraints respect to the open BC.

The following properties of subadditivity and superadditivity on
$W_{LA}$ hold:
\begin{lemma}
\label{lemma:BC}
If $\<C\><\infty$, 
\begin{enumerate}
\item $W^{\#}_{LA}$ is subadditive in $L$ for every fixed $A\in\bt{N}$,
\item $W^{\OBC}_{LA}$ is superadditive in $A$ for every fixed $L\in\bt{N}$,
\item $W^{\CBC}_{LA}$ is subadditive in $A$ for every fixed $L\in\bt{N}$,
\end{enumerate}
respect to a translation on the probability space.
\end{lemma}
\proof
As in the proof of lemma \ref{lemma:OBC<CBC}, the dissipation
on a region $\Lambda(L,A)$ is less or equal to the dissipation on the 
same region on which we impose a greater number of constraints on the 
test potentials.

To prove point (i),  we note that if
$\Lambda(L,A)=\Lambda_1(L_1,A)\bigcup\Lambda_2(L_2,A)$ with $L=L_1+L_2$, then
\[
W^{\#}_{\Lambda(L,A)}\leq W^{\#}_{\Lambda_1(L_1,A)}+W^{\#}_{\Lambda_2(L_2,A)},
\]
since to get the r.h.s.\ we need to impose the constraint that the
potential of the sites with longitudinal coordinate $L_1$ be $v_0L_1$.  
Since the random variables are independent and identically distributed
we can introduce the translation operator in the longitudinal direction
$\tau_l$ on $W^{\#}_{LA}\equiv W^{\#}_{\Lambda(L,A)}$ 
\[
W^{\#}_{L_1+L_2,A}\leq W^{\#}_{L_1A}+W^{\#}_{L_2A}\circ \tau_l^{L_1},
\]
from which the subadditivity of $W^{\#}_{LA}$ respect to $L$ for fixed 
$A\in\bt{N}$.
It is clear that this relation is valid independently of the
contraints, i.e.\ of the BC.

Point (iii) is proven in a similar way, imposing as a constraint the fact that
the potential on the sites of the adjacent boundaries of the regions
$\Lambda'_1(L,A_1)$ e $\Lambda'_2(L,A_2)$, such that
$\Lambda(L,A)=\Lambda'_1(L,A_1)\bigcup\Lambda'_2(L,A_2)$ with $A=A_1+A_2$, 
are proportional to the longitudinal coordinate.   

Finally, to prove point (ii), we observe that the region $\Lambda(L,A)$ with
open BC can be obtained connecting with resistors the nearest-neighbor sites
of the adjacent boundaries of two regions $\Lambda'_1(L,A_1)$ and
$\Lambda'_2(L,A_2)$, also with open BC.  To do this, we must take
$\Lambda(L,A)=\Lambda'_1(L,A_1)\bigcup\Lambda'_2(L,A_2)$ with $A=A_1+A_2$.  
On the other hand, connecting two sites with a resistor
implies passing from a conductance $C=0$ to a conductance
$C\geq 0$.  In this passage the dissipation cannot diminish, i.e.
\[
W^{\OBC}_{\Lambda'_1(L,A)}\geq W^{\OBC}_{\Lambda'_1(L,A_1)}
                                +W^{\OBC}_{\Lambda'_2(L,A_2)}.
\]
Introducing the translation operator in the transverse direction
$\tau_t$, we can write
\[
W^{\OBC}_{L,A_1+A_2}\geq W^{\OBC}_{LA_1}
                          +W^{\OBC}_{LA_2}\circ \tau_t^{A_1}.
\]
from which the superadditivity $W^{\OBC}_{LA}$ respect to $A$ for fixed 
$L\in\bt{N}$.

To complete the proof, we only need to observe that if
the expectation value $\<C\>$ is finite,  the 
expectation values $\ds{\frac{\bt{E}(W_{LA})}{LA}}$ are also finite for
every $L,A\in\bt{N}$.

From lemmas \ref{lemma:BC} and \ref{lemma:additivity}, 
we have the following 
\begin{lemma}
\label{lemma:E(w)}
If $\<C\><\infty$ 
\begin{enumerate}
\item $\ds{\lim_{L\to\infty}\frac{1}{L}}\bt{E}(W^{\#}_{LA})
       =\ds{\inf_L\frac{1}{L}}\bt{E}(W^{\#}_{LA})$,
\item $\ds{\lim_{A\to\infty}\frac{1}{A}}\bt{E}(W^{\OBC}_{LA})
       =\ds{\sup_A\frac{1}{A}}\bt{E}(W^{\OBC}_{LA})$,
\item $\ds{\lim_{A\to\infty}\frac{1}{A}}\bt{E}(W^{\CBC}_{LA})
       =\ds{\inf_A\frac{1}{A}}\bt{E}(W^{\CBC}_{LA})$.
\end{enumerate}	 
\end{lemma}

\subsection{Convergence in the mean}
\label{sec:l.i.m.}

In this section we prove the convergence in the mean (mean square convergence)
\cite{Shiryayev} of $\ds{\frac{1}{LA}W_{LA}}$ as $L\to\infty$ and $A\to\infty$.
For closed BC we demonstrate the convergence independently of the order
of the limits on $L$ and $A$.  For open BC, the result depends on the order
of the limits.  We shall divide the proof in a few preliminary propositions.

\begin{proposition}
\label{prop:E(W_LA)}
If $\<C\><\infty$ the limits
\begin{enumerate}
\item 
$\ds{\lim_{A\to\infty}\frac{\bt{E}(W^{\CBC}_{LA})}{A}}
 =\ds{\inf_A\frac{\bt{E}(W^{\CBC}_{LA})}{A}}\equiv v^2_0L^2\bar g^{\CBC}_L$
\item
$\ds{\lim_{A\to\infty}\frac{\bt{E}(W^{\OBC}_{LA})}{A}}
 =\ds{\sup_A\frac{\bt{E}(W^{\OBC}_{LA})}{A}}\equiv v^2_0L^2\bar g^{\OBC}_L$
\item 
$\ds{\lim_{L\to\infty}\frac{\bt{E}(W^{\CBC,\OBC}_{LA})}{L}}
 =\ds{\inf_L\frac{\bt{E}(W^{\CBC,\OBC}_{LA})}{L}}
\equiv v^2_0\bar c^{\CBC,\OBC}_A$
\end{enumerate}
exist and are finite, with $\bar g^{\OBC}_L \leq \bar g^{\CBC}_L$,
for every $L,A\in\bt{N}$.
\end{proposition}
\proof
We consider first the case of closed BC.
Being the sequence $\bt{E}(W^{\CBC}_{LA})$ bounded from below by zero,
by lemma \ref{lemma:E(w)} the limit
\[
\lim_{A\to\infty}\frac{\bt{E}(W^{\CBC}_{LA})}{A}
               = \inf_{A}\frac{\bt{E}(W^{\CBC}_{LA})}{A} 
          \equiv v^2_0L^2\bar g^{\CBC}_L
\]
exists finite for every $L\in\bt{N}$.  For open BC, by lemma 
\ref{lemma:OBC<CBC}, 
\[
\frac{\bt{E}(W^{\OBC}_{LA})}{A}\leq \frac{\bt{E}(W^{\CBC}_{LA})}{A}.
\]
for every $A\in\bt{N}$, hence passing to the limit $A\to\infty$, 
by lemma \ref{lemma:E(w)} the limit
\[
\lim_{A\to\infty}\frac{\bt{E}(W^{\OBC}_{LA})}{A}
               =\sup_A\frac{\bt{E}(W^{\OBC}_{LA})}{A}
          \equiv v^2_0L^2\bar g^{\OBC}_L 
\]
exists finite, with $\bar g^{\OBC}_L \leq \bar g^{\CBC}_L$,
for every $L\in\bt{N}$.  Point (iii) is proven along similar lines
using the subadditivity in $L$ of $\bt{E}(W_{LA})$, valid independently of
the BC.

\begin{proposition}
\label{prop:c-g}
If $\<C\><\infty$ the limits
\begin{enumerate}
\item 
$\ds{\lim_{A\to\infty}\frac{\bar c^{\CBC}_A}{A}} 
 =\ds{\inf_A\frac{\bar c^{\CBC}_A}{A}}\equiv \bar c^{\CBC}$, 
\item 
$\ds{\lim_{A\to\infty}\frac{\bar c^{\OBC}_A}{A}} 
 =\ds{\sup_A\frac{\bar c^{\OBC}_A}{A}}\equiv \bar c^{\OBC}$, 
\item 
$\ds{\lim_{L\to\infty}}L\bar g^{\CBC,\OBC}_L
 =\ds{\inf_L}L\bar g^{\CBC,\OBC}_L\equiv \bar g^{\CBC,\OBC}$, 
\end{enumerate}
exist and are finite, with $\bar c^{\OBC}\leq \bar c^{\CBC}$.
\end{proposition}
\proof
By lemma \ref{lemma:BC} and proposition \ref{prop:E(W_LA)}, 
it is easily proven that $\bar c^{\CBC}_A$ and $\bar c^{\OBC}_A$
are subadditive and superadditive sequences in $A$, respectively.  
Then, by lemma \ref{lemma:additivity}, being 
$\bar c^{\CBC}_A$ bounded from below by zero, the limit
\[
\lim_{A\to\infty}\frac{\bar c^{\CBC}_A}{A}
               = \inf_{A}\frac{\bar c^{\CBC}_A}{A}
                 \equiv \bar c^{\CBC}
\]
exists finite.  Moreover, by lemma \ref{lemma:OBC<CBC} and proposition
\ref{prop:E(W_LA)}, 
$\bar c^{\OBC}_A\leq \bar c^{\CBC}_A$ for every $A\in\bt{N}$. 
Then, passing to the limit for $A\to\infty$, the limit
\[
\lim_{A\to\infty}\frac{\bar c^{\OBC}_A}{A}
               = \sup_{A}\frac{\bar c^{\OBC}_A}{A} 
               \equiv \bar c^{\OBC}
\]
exists finite, with $\bar c^{\OBC}\leq \bar c^{\CBC}$.
Point (iii) is similarly proven.

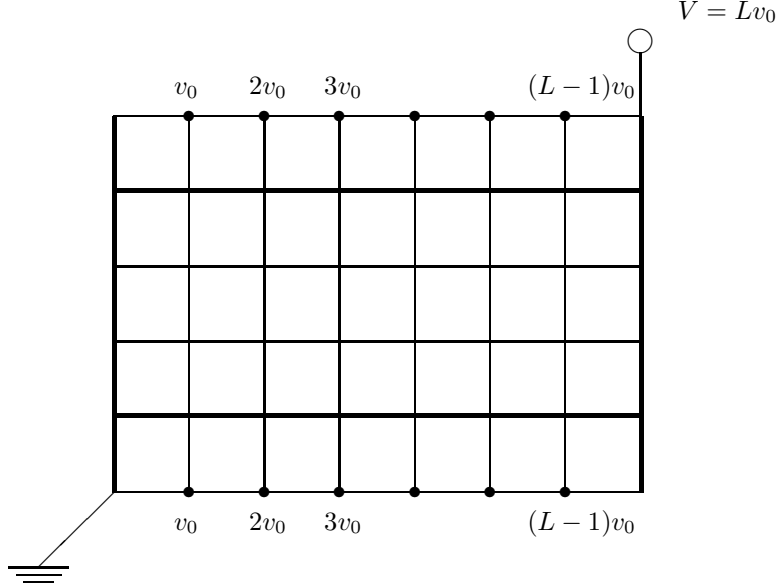
\begin{figure}
\begin{picture}(100,80)(-20,0)
\raggedright
\thinlines
\multiput(10,10)(0,10){6}{\line(1,0){70}}
\multiput(10,10)(10,0){8}{\line(0,1){50}}
\put(10,10){\line(-1,-1){10}}
\put(-4,0){\line(1,0){8}}
\put(-3,-1){\line(1,0){6}}
\put(-2,-2){\line(1,0){4}}
\put(80,60){\line(0,1){8.5}}
\put(80,70){\circle{3}}
\put(85,73){$V=Lv_0$}
\multiput(20,10)(0,50){2}{\circle*{1.5}}
\multiput(18,5)(0,58){2}{$v_0$}
\multiput(30,10)(0,50){2}{\circle*{1.5}}
\multiput(28,5)(0,58){2}{$2v_0$}
\multiput(40,10)(0,50){2}{\circle*{1.5}}
\multiput(38,5)(0,58){2}{$3v_0$}
\multiput(50,10)(0,50){2}{\circle*{1.5}}
\multiput(60,10)(0,50){2}{\circle*{1.5}}
\multiput(70,10)(0,50){2}{\circle*{1.5}}
\multiput(65,5)(0,58){2}{$(L-1)v_0$}
\thicklines
\multiput(10,10)(70,0){2}{\line(0,1){50}}
\multiput(10.3,10)(70,0){2}{\line(0,1){50}}
\multiput(10,20)(0,30){2}{\line(1,0){70}}
\multiput(10,20.3)(0,30){2}{\line(1,0){70}}
\end{picture}
\caption{Comparison of the network of dimensions $(L,A)$ with closed BC
with the network of dimensions $(L,A-2)$ with open BC.}
\label{fig:correction}
\end{figure}

\begin{proposition}
\label{prop:CBC=OBC-A}
If $\<C\><\infty$ then
\begin{equation}
\lim_{A\to\infty}\frac{\bt{E}(W^{\OBC}_{LA})}{A}
		=\lim_{A\to\infty}\frac{\bt{E}(W^{\CBC}_{LA})}{A}.
\end{equation}
\end{proposition}
\proof
By lemma \ref{lemma:OBC<CBC}, it suffices to show that
\begin{equation}
\lim_{A\to\infty}\frac{\bt{E}(W^{\CBC}_{LA})}{A}
		\leq \lim_{A\to\infty}\frac{\bt{E}(W^{\OBC}_{LA})}{A}.
\label{eq:CBC<OBC}
\end{equation}
To prove this inequality, we consider a network of dimensions $(L,A)$ 
with closed BC, and a network of dimensions $(L,A-2)$ with open BC,
as shown in figure~\ref{fig:correction}.

By the Least Dissipation Principle, cf.~theorem \ref{th:ldp}, 
$W^{\CBC}_{LA}\equiv W^{\CBC}_{\Lambda(L,A)}(C,\bar V)$ 
is obtained by minimizing the functional $\varphi(C,\phi)$ 
respect to all the fields $\phi$ that assume the values 
$v_0,2v_0,\ldots,Lv_0$ on the boundaries parallel to the longitudinal
direction.  Denoting by $V$ the extremal field, we have
\[
W^{\CBC}_{LA}=\sum_{(n,n')\in L_{\Lambda(L,A)}}(V_n-V_{n'})^2C_{nn'}.
\]
The extremal field $V'$ for
$W^{\OBC}_{L,A-2}\equiv W^{\OBC}_{\Lambda'(L,A-2)}(C,\bar V')$ is, in general,
not equal to $V$, i.e. it does not minimize $W^{\CBC}_{LA}$.
Thus we have
\begin{eqnarray*}
W^{\CBC}_{LA}&=&\sum_{(n,n')\in L_{\Lambda'(L,A-2)}}(V_n-V_{n'})^2C_{nn'} \\
               &+&\sum_{(n,n')\in L_{\Lambda(L,A)}\bs L_{\Lambda'(L,A-2)}}
                        (V_n-V_{n'})^2C_{nn'} \\
            &\leq&\sum_{(n,n')\in L_{\Lambda'(L,A-2)}}(V'_n-V'_{n'})^2C_{nn'} \\
               &+&\sum_{\{n\in\Lambda'(L,A-2),\ 
                        n'\not\in \Lambda'(L,A-2):\ |n-n'|=1\}}
                        (V'_n-n'_1v_0)^2C_{nn'} \\
               &+&\sum_{(n,n')\in L_{\Lambda(L,A)\bs\Lambda'(L,A-2)}}
                        (n_1v_0-n'_1v_0)^2C_{nn'}
\end{eqnarray*}
from which 
\begin{equation}
\frac{\bt{E}(W^{\CBC}_{LA})}{A}\leq \frac{\bt{E}(W^{\OBC}_{L,A-2})}{A}
+\frac{\bt{E}(W^{\tr}_{LA})}{A}+\frac{\bt{E}(W^{\bound}_L)}{A},
\label{eq:A<A-2}
\end{equation}
where 
\begin{equation}
\label{eq:bound}
W^{\bound}_L=v^2_0\sum_{k=1}^{2L}C^{\bound}_k
\end{equation}
is the dissipation due to the conductances $C^{\bound}_k,\ k=1,\ldots,2L$ 
which lie on the longitudinal boundaries of the network of dimensions $(L,A)$ 
with closed BC and
\begin{equation}
W^{\tr}_{LA}=W^{\Sup}_{LA}+W^{\Inf}_{LA}
\end{equation}
is the dissipation due to the transverse conductances 
of $L_{\Lambda(L,A)}\bs L_{\Lambda'(L,A-2)}$ that do not lie on the 
equipotential sides, which don't dissipate.  The superscripts
sup and inf indicate the contributions of the superior and inferior part of
$\Lambda(L,A)\bs\Lambda'(L,A-2)$.  We indicate with $C^{\Sup}_k$ and
$C^{\Inf}_k$, $k=1,\ldots,L-1$ these conductances,  and with $V^{\Sup}_k$ 
and $V^{\Inf}_k$, $k=1,\ldots,L-1$, the potentials on the sites on  the
two longitudinal boundaries of the network of dimensions
$(L,A-2)$ with open BC, excluding the sites that lie on the equipotential sides.
Thus, we have
\begin{equation}
W^{\Sup,\Inf}_L=v^2_0\sum_{k=1}^{L-1}(V^{\Sup,\Inf}_k-kv_0)^2C^{\Sup,\Inf}_k.
\end{equation}
Note that transverse term $W^{\tr}_{LA}$ in general can depend both on 
$L$ and $A$, while the boundary term $W^{\bound}_L$ does not depend on $A$, so
it does not contribute to the limit for $A\to \infty$ of equation 
(\ref{eq:A<A-2}).  Thus, we only need to show that 
\begin{equation}
\label{eq:E(W-tr)=0}
\lim_{A\to\infty}\frac{\bt{E}(W^{\tr}_{LA})}{A}=0,
\end{equation}
from which, since
\begin{equation}
\frac{\bt{E}(W^{\OBC}_{L,A-2})}{A}\leq \frac{\bt{E}(W^{\OBC}_{L,A-2})}{A-2}, 
\end{equation}
i.e.~equation (\ref{eq:CBC<OBC}).

To estimate the term $\bt{E}(W^{\tr}_{LA})$, note that by
the Maximum Principle, cf.~proposition \ref{prop:max}, we have
$0\leq V^{\Sup,\Inf}_{k}\leq Lv_0$, for every $k<L$, from which
\begin{equation}
\label{eq:estimate-max-sup}
|V^{\Sup,\Inf}_{k}-kv_0|\leq (L-1)v_0
\end{equation}
for every $k<L$.  Then, 
$W^{\Sup,\Inf}_{LA}\leq (L-1)^2v^2_0\sum_{k=1}^{L-1}C^{\Sup,\Inf}_{k}$
and, being the variables identically distributed, 
\begin{equation}
\bt{E}(W^{\tr}_{LA})\leq 2(L-1)^3v^2_0\<C\>.
\end{equation}
Since this estimate is uniform in $A$, we obtain equation~(\ref{eq:E(W-tr)=0})
and thus the thesis.

We can sum up the preceding propositions in the following:
\begin{theorem}
\label{th:l.i.m.}
If $\<C\><\infty$, the limit
\begin{equation}
\lim_{L,A\to\infty}\frac{\bt{E}(W^{\CBC}_{LA})}{LA}=v^2_0\bar c^{\CBC}
\end{equation}
exist and is finite, independently of the order with which
we take the limits, while
\begin{eqnarray*}
\lim_{A\to\infty}\lim_{L\to\infty}\frac{\bt{E}(W^{\OBC}_{LA})}{LA}
                                              &=&v^2_0\bar c^{\OBC}\\
\lim_{L\to\infty}\lim_{A\to\infty}\frac{\bt{E}(W^{\OBC}_{LA})}{LA}
                                              &=&v^2_0\bar c^{\CBC},
\end{eqnarray*}
with $\bar c^{\OBC}\leq \bar c^{\CBC}$.  
\end{theorem}
\proof
By propositions \ref{prop:E(W_LA)} and \ref{prop:c-g},
\begin{eqnarray*}
\lim_{L\to\infty}\lim_{A\to\infty}\frac{\bt{E}(W^{\CBC,\OBC}_{LA})}{LA}
                                                &=& v^2_0 \bar g^{\CBC,\OBC},\\
\lim_{A\to\infty}\lim_{L\to\infty}\frac{\bt{E}(W^{\CBC,\OBC}_{LA})}{LA}
                                                &=& v^2_0 \bar c^{\CBC,\OBC},
\end{eqnarray*}
so we only need to show that $\bar g^{\OBC}=\bar g^{\CBC}=\bar c^{\CBC}$.
The first equality follows from proposition \ref{prop:CBC=OBC-A}, since
\[
\bar g^{\CBC}_L=\lim_{A\to\infty}\frac{\bt{E}(W^{\CBC}_{LA})}{A}
                 =\lim_{A\to\infty}\frac{\bt{E}(W^{\OBC}_{LA})}{A}
                 =\bar g^{\OBC}_L,
\]
for every $L\in\bt{N}$.  The second equality from the fact that
\[
\bar g^{\CBC}=\inf_L L\bar g^{\CBC}_L 
            =\inf_{L,A}\frac{\bt{E}(W^{\CBC}_{LA})}{LA} 
            =\inf_A\frac{\bar c^{\CBC}_A}{A}=\bar c^{\CBC}.   
\]

\subsection{Almost sure convergence}
\label{sec:l.q.c.}

In this section we prove the main results of this paper. 
We first show that, for closed BC, the dissipation density
$\ds{\frac{W_{LA}}{LA}}$ converges almost surely, as $L,A\to\infty$,
independently of the order of the limits.  This is done by exploiting
general theorems on subadditive sequences \cite{Kingman,BCES}, reported in 
\ref{app:sss}.
We also show that the a.s.\ convergence  holds for both open and closed BC
if we let $A\to\infty$ before $L\to\infty$, i.e. 
we prove the independence of the boundary conditions
for a given order of the limits.  
The novelty of this second result is that the different
behaviour of the dissipation with open and closed BC is exploited 
to prove the a.s.\ convergence,
using classical theorems of probability theory, such as Kolmogorov's 
strong law of large numbers \cite{Shiryayev}, instead of the general 
theorems of \ref{app:sss}.

\subsubsection{Independence of the order of the limits for closed BC}

\begin{theorem} 
\label{th:l.q.c.1}
For closed BC, the limit
\begin{equation}
\lim_{L,A\to\infty}\frac{W^{\CBC}_{LA}}{LA}=v^2_0\bar c^{\CBC}
\end{equation}
exists a.s. and is finite, independently of the order of the limits, where
\begin{equation}
v^2_0\bar c^{\CBC}=\lim_{L,A\to\infty}\frac{\bt{E}(W^{\CBC}_{LA})}{LA}.
\end{equation}
\end{theorem}
\proof
We must show that a.s. the limits
\begin{enumerate}
\item $\ds{\lim_{A\to\infty}\lim_{L\to\infty}\frac{W^{\CBC}_{LA}}{LA}}
                      =v^2_0\bar c^{\CBC}$,
\item $\ds{\lim_{L\to\infty}\lim_{A\to\infty}\frac{W^{\CBC}_{LA}}{LA}}
                      =v^2_0\bar c^{\CBC}$,
\end{enumerate}
exist and are finite.
By lemma \ref{lemma:BC}, $W^{\CBC}_{LA}$ is subadditive in $L$ for every 
fixed $A$, and in $A$ for every fixed $L$, with a translation as a 
measure-preserving transformation on the probability space.  
Hence, by the theorems in \ref{app:sss} and proposition \ref{prop:E(W_LA)}, 
we get that the limits
\begin{eqnarray*}
\lim_{L\to\infty}\frac{W^{\CBC}_{LA}}{LA}&=&
          \inf_L\frac{\bt{E}(W^{\CBC}_{LA})}{LA}=v^2_0\bar c^{\CBC}_A \\
\lim_{A\to\infty}\frac{W^{\CBC}_{LA}}{LA}&=&
          \inf_A\frac{\bt{E}(W^{\CBC}_{LA})}{LA}=v^2_0L\bar g^{\CBC}_L
\end{eqnarray*}
exist a.s. and are finite.  Then, by
proposition \ref{prop:c-g}, the limits
\begin{eqnarray*}
\lim_{A\to\infty}\lim_{L\to\infty}\frac{W^{\CBC}_{LA}}{LA}
                      &=&v^2_0\bar c^{\CBC} \\
\lim_{L\to\infty}\lim_{A\to\infty}\frac{W^{\CBC}_{LA}}{LA}
                      &=&v^2_0\bar g^{\CBC},
\end{eqnarray*}
also exist a.s. and are finite.
The thesis follows from theorem \ref{th:l.i.m.}.

\subsubsection{Independence of the boundary conditions for a given order of
the limits}

\begin{theorem} 
\label{th:l.q.c.2}
Independently of the BC, the limit
\begin{equation}
\lim_{L\to\infty}\lim_{A\to\infty}\frac{W_{LA}}{LA}=v^2_0\bar c^{\CBC}
\end{equation}
exists a.s. and is finite, where
\begin{equation}
v^2_0\bar c^{\CBC}=\lim_{L\to\infty}\lim_{A\to\infty}
                     \frac{\bt{E}(W^{\CBC}_{LA})}{LA}.
\end{equation}
\end{theorem}
\proof
Let us consider an increasing sequence of integers $\{A_p\}$ such that
$A_p\to \infty,\ p\to\infty$.
We fix our attention on a generic element of the sequence $A_p$ such that
$A=N_pA_p+r_p,\ r_p<A_p$.
Let us divide the network of transverse extension $A$ in $N_p$ networks of
extension $A_p$ plus a network of extension $r_p$.  Being $W^{\OBC}_{LA}$ 
a superadditive sequence  in $A$ for every fixed $L\in\bt{N}$, we have
\begin{equation}
\label{eq:W-split-ap}
W^{\OBC}_{LA}\geq \sum_{k=1}^{N_p}W^{\OBC}_{LA_p}\circ\tau^{(k-1)A_p} 
                    + W^{\OBC}_{Lr_p}\circ\tau^{NA_p},
\end{equation}
where $W^{\OBC}_{LA_p}\circ\tau^{(k-1)A_p} ,\ k=1,\ldots,N_p,$ are
independent and identically distributed random variables.  Dividing 
both sides by $A-r_p=N_pA_p$, and letting $A\to\infty$, or equivalently 
$N_p\to\infty$,
\[
\liminf_{A\to\infty}\frac{W^{\OBC}_{LA}}{A}\geq\lim_{N_p\to\infty}
\frac{1}{N_p}\sum_{k=1}^{N_p}\frac{W^{\OBC}_{LA_p}\circ\tau^{(k-1)A_p}}{A_p}. 
\]
Since the expectation value $\ds{\frac{1}{A_p}}\bt{E}(W^{\OBC}_{LA_p})$ is 
finite we can apply Kolmogorov's strong law of large numbers \cite{Shiryayev}
to state that $\bt{P}(\ct{N}^{\OBC}_p)=0$, where
\[
\ct{N}^{\OBC}_p=\left\{\liminf_{A\to\infty}\frac{W^{\OBC}_{LA}(\omega)}{A}
                    < \frac{\bt{E}(W^{\OBC}_{LA_p})}{A_p}\right\}.
\]
Using the subadditivity of the measure, we obtain 
\[
\bt{P}\left(\bigcap_{p=1}^{\infty}\bar{\ct{N}}^{\OBC}_p\right)=\bt{P}\left\{ 
              \liminf_{A\to\infty}\frac{W^{\OBC}_{LA}(\omega)}{A}
         \geq \frac{\bt{E}(W^{\OBC}_{LA_p})}{A_p},\ 
\forall\ p\in\bt{N}\right\}=1,
\]
and, by proposition \ref{prop:E(W_LA)},
\[
\bt{P}\left\{\liminf_{A\to\infty}\frac{W^{\OBC}_{LA}(\omega)}{A}
          \geq v^2_0L^2\bar g^{\OBC}_L\right\}=1.
\]
For closed BC we can use similar arguments 
using the subadditivity of $W^{\CBC}_{LA}$ instead of the superadditivity
of $W^{\OBC}_{LA}$, and we obtain 
\[
\bt{P}\left\{\limsup_{A\to\infty}\frac{W^{\CBC}_{LA}(\omega)}{A}
                     \leq  v^2_0L^2\bar g^{\CBC}_L\right\}=1.
\]
Then, by lemma \ref{lemma:OBC<CBC} and proposition \ref{prop:CBC=OBC-A}, 
which implies $\bar g^{\CBC}_L=\bar g^{\OBC}_L\equiv\bar g_L$, 
we find that the limit 
\begin{equation}
\lim_{A\to\infty}\frac{W^{\CBC,\OBC}_{LA}(\omega)}{A}=v^2_0L^2\bar g_L
\label{eq:lim-A-W}
\end{equation}
exists finite for every
$\omega\in\Omega\bs(\ct{N}^{\CBC,\OBC}_{\Sup}
\bigcap\ct{N}^{\CBC,\OBC}_{\Inf})$, where
we have introduced the following sets of zero measure
\begin{eqnarray*}
\ds{\ct{N}^{\CBC}_{\Sup}=\left\{\limsup_{A\to\infty}
    \frac{W^{\CBC}_{LA}(\omega)}{A}>v^2_0L^2\bar g_L \right\}}, \\
\ds{\ct{N}^{\CBC}_{\Inf}=\left\{\liminf_{A\to\infty}
    \frac{W^{\CBC}_{LA}(\omega)}{A}<v^2_0L^2\bar g_L \right\}}, \\
\ds{\ct{N}^{\OBC}_{\Sup}=\left\{\limsup_{A\to\infty}
    \frac{W^{\OBC}_{LA}(\omega)}{A}>v^2_0L^2\bar g_L \right\}}, \\
\ds{\ct{N}^{\OBC}_{\Inf}=\left\{\liminf_{A\to\infty}
    \frac{W^{\OBC}_{LA}(\omega)}{A}<v^2_0L^2\bar g_L \right\}}.
\end{eqnarray*}
Since $\bt{P}(\ct{N}^{\CBC,\OBC}_{\Sup}\bigcap\ct{N}^{\CBC,\OBC}_{\rm inf})=0$,
the limit (\ref{eq:lim-A-W}) exist a.s. and is finite both for open and 
closed BC. The thesis follows from theorem \ref{th:l.i.m.}
and proposition \ref{prop:c-g}.


\appendix

\section{Theorems on subaddivite and superadditive sequences}
\label{app:sss}

\begin{theorem}[Kingman \cite{Kingman}]
\label{th:Kingman}
Let $\{\xi_n\},\ n\in\bt{N}$, be a subadditive sequence such that
$\bt{E}(\xi_n)\geq -An$ for some positive constant $A$.  Then, the limit
\begin{equation}
\xi=\lim_{n\to\infty}\frac{\xi_n}{n}
\end{equation}
exist finite almost surely and in the mean and $\bt{E}(\xi)=\gamma$, where
\begin{equation}
\gamma=\inf_n\frac{\bt{E}(\xi_n)}{n}.
\end{equation}
Moreover, the limit $\xi$ can be represented as
\begin{equation}
\xi=\lim_{n\to\infty}\frac{\bt{E}(\xi_n|\ct{A})}{n},
\end{equation}
where $\ct{A}$ is the $\sigma$-algebra of the events invariant under the 
measure-preserving transformation on the probability space 
which defines the subadditive process.  
In particular, if $\ct{A}$ contains only events
of probability 0 or 1, then $\xi=\gamma$.
\end{theorem}

The limit variable $\xi$ is degenerate only in particular cases.
For example, if $\xi_n=F_n(\eta_1,\eta_2,\ldots)$,
where $\eta_1,\eta_2,\ldots$ are independent and identically distributed 
variables, then by the zero-one law the $\sigma$-algebra $\ct{A}$ 
is trivial and $\xi=\gamma$ \cite{Kingman}.  Indeed, this is the case for the
problem at hand, where the sequence $\xi_n$ 
is given by the dissipation as a function of the volume of the region $\Lambda$
and the variables $\eta_1,\eta_2,\ldots$ are the dissipations associated to the
subvolumes of $\Lambda$.

Actually, in the case that the transformation is a translation
one can show directly that $\xi$ is degenerate, i.e. we have the 
following\footnote{The results obtained for superadditive sequences 
can be easily reformulated for subadditive sequences, and viceversa.}
\begin{theorem}[Bellisard {\em et al}.\ \cite{BCES}]
\label{th:BCES}
Let $\{\xi_n\},\ n\in\bt{N}$, be a superadditive sequence with 
measure-preserving translation $\tau$ on the probability space, such that
$\bt{E}(\xi_n)\leq An$ for some positive constant $A$.  Then, the limit
\begin{equation}
\xi=\lim_{n\to\infty}\frac{\xi_n}{n}
\end{equation}
exist finite almost surely and in the mean and $\xi=\gamma$, where
\begin{equation}
\gamma=\sup_n\frac{\bt{E}(\xi_n)}{n}.
\end{equation}
\end{theorem}

We also note the following lemma \cite{Kingman}:
\begin{lemma}
\label{lemma:additivity}
Let $\{a_n\},\ n\in\bt{N}$ be a numeric sequence.  Then
\begin{enumerate}
\item $\ds{\lim_{n\to\infty}\frac{a_n}{n}}=\ds{\inf_n\frac{a_n}{n}}$ 
      if $a_n$ is subadditive,
\item $\ds{\lim_{n\to\infty}\frac{a_n}{n}}=\ds{\sup_n\frac{a_n}{n}}$ 
      if $a_n$ is superadditive.
\end{enumerate}
\end{lemma}

\end{document}